\documentclass[floatfix, aip, rsi,%
 amsmath,amssymb,
reprint,%
usenames,dvipsnames
]{revtex4-1}

\usepackage{graphicx}% Include figure files
\usepackage{dcolumn}% Align table columns on decimal point
\usepackage{bm}% bold math
\usepackage{gensymb}
\usepackage{soul,xcolor}

\usepackage{ulem}
\newcommand{\sss}[1]{%
    \renewcommand{\ULthickness}{0.8pt}%
       \sout{#1}%
    \renewcommand{\ULthickness}{.4pt}% Resetting to ulem default
}
\setstcolor{NavyBlue}

\begin{document}

\preprint{AIP/123-QED}

\title[]{Heterogeneous Seeded Molecular Dynamics \\ as a Tool to Probe the Ice Nucleating Ability of Crystalline Surfaces}
%\thanks{Footnote to title of article.}

\author{Philipp Pedevilla}
\affiliation{Thomas Young Centre, London Centre for Nanotechnology 
and Department of Physics and Astronomy, University College London, Gower Street London, London, WC1E 6BT, UK}
\author{Martin Fitzner}
\affiliation{Thomas Young Centre, London Centre for Nanotechnology 
and Department of Physics and Astronomy, University College London, Gower Street London, London, WC1E 6BT, UK}
\author{Gabriele C. Sosso}
\affiliation{Department of Chemistry and Centre for Scientific Computing, University of Warwick, Gibbet Hill Road, Coventry, CV4 7AL, UK}
\email{G.Sosso@warwick.ac.uk}
\author{Angelos Michaelides}
\affiliation{Thomas Young Centre, London Centre for Nanotechnology
and Department of Physics and Astronomy, University College London, Gower Street London, London, WC1E 6BT, UK}
\email{angelos.michaelides@ucl.ac.uk}

\date{\today}% It is always \today, today,

\begin{abstract}

Ice nucleation plays a significant role in a large number of natural and technological processes, but it is challenging
to investigate experimentally because of the small time (ns) and short length scales (nm) involved. On the other hand,
conventional molecular simulations struggle to cope with the relatively long timescale required for 
critical ice nuclei to form. One way to tackle this issue
is to take advantage of free energy or path sampling techniques.  Unfortunately,
these are computationally costly.  Seeded molecular dynamics is a much less demanding alternative that has been
successfully applied already to study the homogeneous freezing of water.  However, in the case of {\it heterogeneous}
ice nucleation, nature's favourite route to form ice,  an array of suitable interfaces between the ice seeds and the
substrate of interest has to be built - and this is no trivial task.  In this paper, we present a Heterogeneous SEEDing
approach (HSEED) which harnesses a random structure search framework to tackle the ice-substrate challenge, thus
enabling seeded molecular dynamics simulations of heterogeneous ice nucleation on crystalline surfaces.  We validate the
HSEED framework by investigating the nucleation of ice on: (i) model crystalline surfaces, using the coarse-grained mW
model; and (ii) cholesterol crystals, employing the fully atomistic TIP4P/Ice water model. We show that the
HSEED technique yields results in excellent agreement with both metadynamics and forward flux sampling simulations.
Because of its computational efficiency, the HSEED method allows one to rapidly assess the ice nucleation
ability of whole libraries of crystalline substrates - a long-awaited computational development in e.g. atmospheric
science.

\end{abstract}

%\keywords{}%Use showkeys class option if keyword
\maketitle

\section{\label{sec:INTRO}Introduction}

The nucleation of ice is the microscopic phenomenon at the heart of one of the most important phase transitions on earth, that is the freezing of water.
For instance, organisms living in cold conditions need to prevent ice formation in their
cells to stay alive~\cite{Mazur_Science_1970_Cryobiology, Lintunen_SciRep_2013_IN-trees}. The formation of ice is of
relevance to atmospheric science as well: the amount of ice in clouds represents a crucial parameter in climate
modelling and it also determines the extent to which solar radiation penetrates into the
atmosphere~\cite{Pratt_NatGeosci_2009_bio-in-clouds,Murray_ChemSocRev_2012_INPinCloudDroplets,
Bartels_Nature_2013_ice-chemistry}. In addition, a thorough understanding of how water freezes into ice is key to
industrial applications such as cryogenic technologies~\cite{Tam_JACS_2009_cryotherapy}, fossil fuel
extraction~\cite{Koh_ChemSocRev_2002_gas-hydrates}, aviation~\cite{Murray_Fuel_2011_INFuel} and many more.

Despite its importance, it is challenging to characterise ice nucleation experimentally, due to the short time scale
involved (of the order of nanoseconds), the small size of the ice nuclei (typically nanometres) and the stochastic
nature of nucleation events. Molecular simulations can in principle be used to learn more about the formation of ice
{\it in silico}, and indeed they have recently been extensively used to get microscopic insight into the nucleation
process (see e.g.
Refs.~\citenum{Sosso_ChemRev_2016_IceReview,Zielke_JPCB_2015_AgI,Zhang_JCP_2014_surface-structure-INA,Reinhardt_JCP_2014_surface-INA,Guillaume_JCP_2014_AgI,Kiselev_Science_2016_Fsp100Nuc,bi_enhanced_2017,lupi_role_2017}).
However, the time needed for the ice nuclei to become ``critical'', that is large enough to overcome the free energy
barrier preventing them to grow into actual ice crystals, is typically several orders of magnitude longer than the time
scale accessible to e.g. classical molecular dynamics (MD) simulations~\cite{Sosso_ChemRev_2016_IceReview}. Direct
observation of homogeneous water freezing can be achieved via brute force coarse-grained simulations (see e.g.
Ref.~\citenum{Hudait_JACS_2016_IcePolymorphClouds}), most prominently by taking advantage of the mW model of
water~\cite{Molinero_JPCB_2008_mW}. However, in order to nucleate ice from supercooled liquid water using fully
atomistic water models, enhanced sampling methods have to be employed. Various options are available: free energy based
methods such as umbrella sampling~\cite{Torrie_JComputPhys_1977_UmbrellaS, Warmflash_JCP_2007_UmbrellaS,
TenWolde_JCP_1996_NucRateLJ, Auer_Nature_2001_ColloidNucRate, Radhakrishnan_JACS_2003_IhNucleation} and
metadynamics~\cite{Laio_PNAS_2002_MetaDynamics, Trudu_PRL_2006_MetaDynLJ, Salvalaglio_2017_JCP_WellTemperedMetaDyn}, as
well as path sampling methods such as transition path sampling~\cite{Bolhuis_AnnuRevPhysChem_2002_TPS,
Lechner_PhysRevLett_2011_IceNuc} and forward flux sampling (FFS)~\cite{Valeriani_JCP_2005_NaClNuc,
Filion_JCP_2010_USvsFFS, Li_PCCP_2011_HomoNuc, Akbari_2015_PNAS_FFS-IN-homo, Sosso_JPCL_2016_FFS-kaolinite}.

All of these methods are computationally expensive. As an extreme example, the especially thorough investigation of
homogeneous water freezing carried out by Haji-Akbari and Debenedetti~\cite{Akbari_2015_PNAS_FFS-IN-homo} required ca.
21,000,000 CPU hours. This is the reason why, even by taking advantage of state-of-the-art enhanced sampling techniques,
computer simulations of ice nucleation are more often than not performed only at very strong supercooling
($T_\mathrm{m}-T=\Delta T_\mathrm{S}\sim$ 40 K, where $T_\mathrm{m}$ stands for the melting temperature of ice). This is
sub-optimal, as making a connection between simulations and experiments requires to collect results at different
temperatures - mild supercooling included. In fact, the absolute values of thermodynamic and kinetic properties such as
the critical nucleus size $N_\mathrm{C}^{*}$ and the ice nucleation rate~\cite{kalikali}, respectively, are exceedingly
sensitive to a number of computational details~\cite{Sosso_ChemRev_2016_IceReview}, chiefly the accuracy of the water
model employed~\cite{Akbari_2015_PNAS_FFS-IN-homo}, so that a single absolute value of e.g. the nucleation rate at a
given supercooling is of little practical relevance.

Seeded MD (see e.g. Refs.~\citenum{Espinosa_JChemPhys_2016_Homo-Seeding}) represents one way to overcome these limitations, and involves
the monitoring in time of a collection of MD trajectories at different temperatures, where ice nuclei of different size (and possibly shape) have
been inserted into supercooled liquid water {\it beforehand}. At a given temperature, these ice ``seeds'', i.e. nuclei smaller or larger than $N_\mathrm{C}^{*}$, 
would dissolve or grow respectively, 
thus allowing one to pinpoint the critical nucleus size itself. This approach is computationally very efficient, and thus applicable to mild supercooling. On the other hand, it does not provide direct information about the actual nucleation mechanism 
(how exactly water molecules come to form a critical ice nucleus), and it relies on the assumption that we can guess {\it a priori} the
structural properties (shape, crystalline polytype...) of the ice seeds. Moreover, to obtain quantities of interest to 
experimentalists such as the ice nucleation rate, a number of additional parameters such as the interfacial free energy between water and ice have to be calculated
according to classical nucleation theory (CNT~\cite{kalikali}). 

The success of the seeding technique is due to the fact that the shape and the composition of the
crystalline seeds is often well known {\it a priori}. In the case of ice, cubic ice (I$_\mathrm{c}$) and
hexagonal ice (I$_\mathrm{h}$) are the two potential candidates, but mixtures of the two (a crystalline phase known as stacking disordered ice, I$_\mathrm{sd}$) have
also been reported in both experiments~\cite{Malkin_PCCP_2015_StackingDisoreredIce} and simulations~\cite{C1CP22022E}. Moreover,
CNT assumes that the seeds have to be spherical, so as to minimise the extent of the crystal/nucleus interface. This approximation
is not necessarily robust at strong supercooling~\cite{Sosso_ChemRev_2016_IceReview}. However, 
Zaragoza {\it et al.}~\cite{Zaragoza_JCP_2015_IcIhSphericalNuclei} found that even cubically shaped ice seeds
reconstruct into a spherical morphology within a few ns of MD simulations - at mild and strong supercooling alike.
In addition, I$_\mathrm{c}$ and I$_\mathrm{h}$ seeds yielded the same nucleation rate~\cite{Zaragoza_JCP_2015_IcIhSphericalNuclei}, 
thus making seeded MD simulations a relatively straightforward computational technique to probe the homogeneous freezing of water.

In nature as well as in the laboratory, though, water almost always freezes heterogeneously, i.e.
thanks to the presence of impurities that promote the kinetics of ice formation.
Very diverse materials can facilitate the heterogeneous nucleation of ice, from mineral dust to birch 
pollen~\cite{Murray_ChemSocRev_2012_INPinCloudDroplets}, and what is it that make these substances capable of
boosting the kinetics of water freezing is still not fully understood~\cite{Sosso_ChemRev_2016_IceReview}.
Simulations have provided useful insight into the molecular details of ice formation on a variety of different 
compounds, in most cases by taking advantage of the coarse-grained mW model for water~\cite{Moore_PhysChemChemPhys_2010_nanopore,Lupi_JACS_2014_Ccurvessize,Zhang_JCP_2014_surface-structure-INA,Reinhardt_JCP_2014_surface-INA,Cox_JCP_2015_layers,Cox_JCP_2015_hydrophilicity,Fitzner_JACS_2015_HeteroNuc-LJ,
Bi_JPhysChemC_2016_CrystallinityHydrophilicity,Lupo_JChemPhys_2016_interfacialwater-twostep}. 
However, addressing the freezing
of water at complex interfaces, such as minerals, organic crystals, and biological matter, requires the use of fully atomistic water models in order
to capture the subtleties of the hydrogen bond network in the proximity of the impurity. 
FFS simulations have been recently used to compute the heterogeneous ice nucleation rate on the clay mineral kaolinite~\cite{Sosso_JPCL_2016_FFS-kaolinite} using an atomistic water model,
but the substantial computational cost limited the investigation to a single crystalline surface at a specific (strong) supercooling.
This is why it would be desirable to extend the capabilities of seeded MD to the study of heterogeneous ice nucleation. 

In this work we present a Heterogeneous SEEDing
approach (HSEED) which harnesses a random structure search (RSS) algorithm to explore the configurational space of different ice seeds
sitting on arbitrary crystalline surfaces, thus enabling seeded MD simulations of heterogeneous ice nucleation. 
While the HSEED method does not offer the same level of detail and accuracy of free energy- and path sampling-based methods, it is
orders of magnitude faster, thus allowing one to investigate different substrates at different temperatures.
We demonstrate the capabilities of the HSEED method by validating its outcomes against: (i) free energy (metadynamics) simulations of mW water freezing on top
of Lennard-Jones (LJ) crystals; and (ii) path sampling (FFS) simulations of a fully atomistic water model on cholesterol (CHL) crystals. 
The HSEED framework consistently pinpoints the same morphologies (in terms of e.g. structure, orientation, ice polytype...) of the ice seeds 
we observe in our metadynamics (FFS) simulations of water freezing on LJ (CHL) crystals. Importantly, we show that
the method allows one to obtain qualitative estimates of the critical ice nucleus size.
Assuming the validity of CNT, one can thus calculate the ice nucleation
rate by comparing the heterogeneous critical nucleus size with its homogeneous counterpart -- albeit this comparison has to be treated with great 
care (as discussed in section~\ref{sec:HSEED}). Most importantly, the HSEED method can be used to rapidly screen the ice nucleating ability of whole libraries of 
crystalline materials and surfaces, allowing one to extract invaluable trends of practical interest for experiments and applications.

The remainder of this paper is structured as follows. The HSEED framework is illustrated in Section~\ref{sec:METHODS}, and we present in Section~\ref{sec:RESULTS} the results
of the method applied to mW water freezing into ice on LJ crystals (Section~\ref{sec:MW}) and to the formation of ice
(from TIP4P/Ice water) on CHL
crystals (Section~\ref{sec:TIP4P}). A discussion of the main outcomes of this work and of the potential future applications of the HSEED method can be found in
Section~\ref{sec:Discussion}.

\section{\label{sec:METHODS}Computational Methods}

\subsection{\label{sec:HSEED} Heterogeneous Seeded Molecular Dynamics}

\begin{figure}
\begin{centering}
\centerline{\includegraphics[width=8cm]{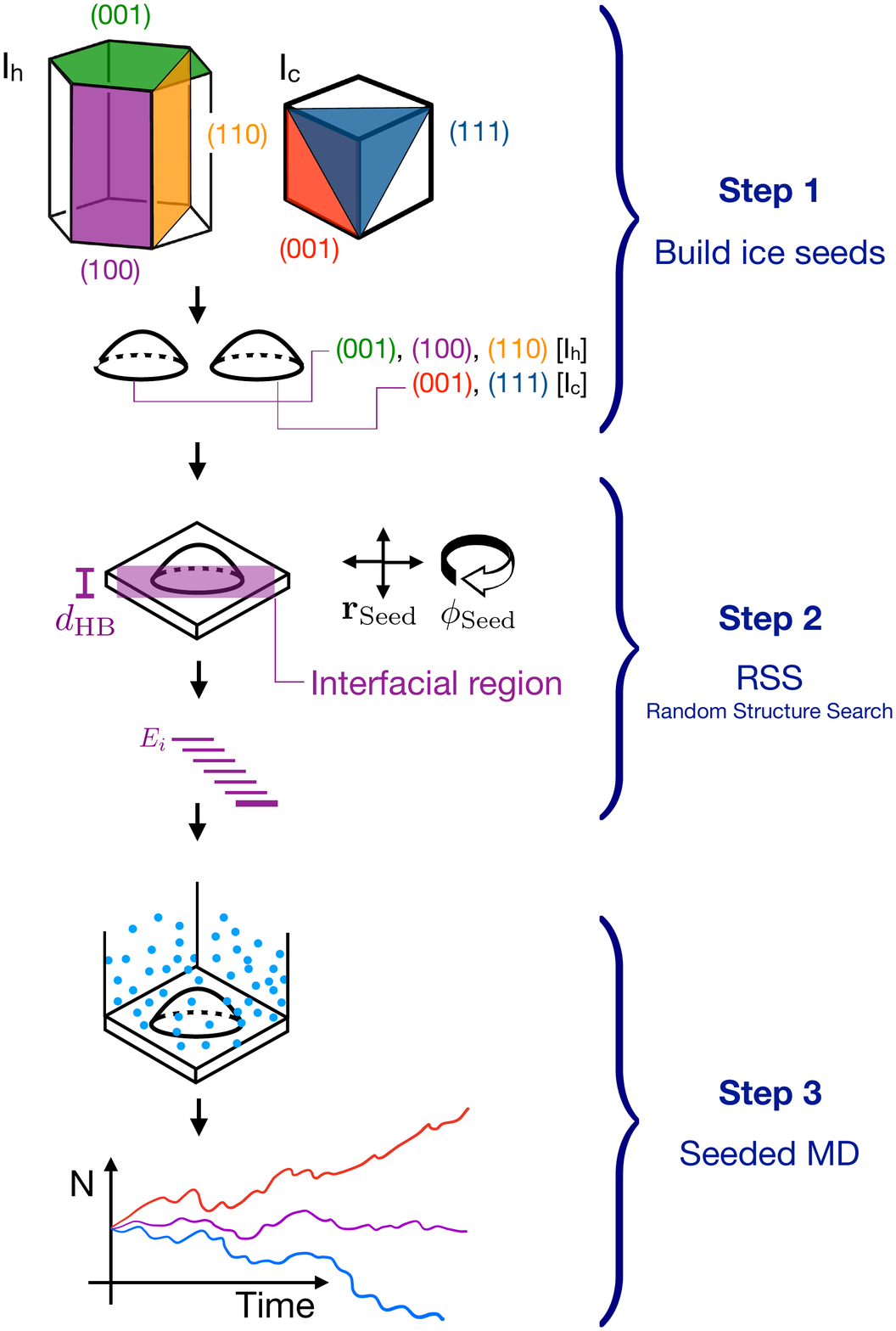}}
\par\end{centering}
\protect\caption{Flowchart of the HSEED method. {\bf Step 1}: From the bulk phases of I$_\mathrm{h}$ and I$_\mathrm{c}$ (I$_\mathrm{sd}$ could also be considered when building large enough
seeds), spherical caps of a certain size, exposing a selection of low-Miller-index
surfaces, are built. {\bf Step 2}: By means of a random  structure search (RSS) algorithm, different locations/orientations and different
configurations of the ice seed-substrate interfacial region are explored. A geometry optimisation of the interfacial region of each one of these configurations is  
then performed, and the resulting structures are ranked according to their potential energies. {\bf Step 3}: The ``best'' candidates, selected following the two criteria detailed
in Sec.~\ref{sec:HSEED}, are solvated in liquid water and then used as starting points for seeded MD simulations.}
\label{FIG_1}
\end{figure}

The key step of the seeded MD framework~\cite{Espinosa_JChemPhys_2016_Homo-Seeding} is the choice/construction of the
crystalline seeds. As discussed in the previous section, this is a relatively straightforward task when dealing with homogeneous water freezing -- but
it becomes more challenging in the heterogeneous nucleation scenario, for the reasons outlined below.

\subsubsection{The shape of the nuclei}

Heterogeneous CNT relies on the assumption that crystalline nuclei of any given size are shaped as spherical caps.
This is a reasonable approximation for large nuclei, where such a shape would minimise the interfacial free energy between the ice seeds and the
supercooled liquid phase. At strong supercooling, however, where the critical nuclei can contain of the order of 10$^2$ molecules only, the templating
effect of the substrate could lead to very anisotropic seeds. In fact, a large body of work has shown the emergence of unique water/ice-like 
structures forming on crystalline surfaces~\cite{Bjorneholme_ChemRev_2016_WaterAtInterfaces}: predicting the topology of these water clusters and/or 
ice-like structures on a given substrate is a challenging task.
Cabriolu and Li~\cite{Cabriolu_PRE_2015_hetCNTwithmW} found that ice nuclei
of mW water nucleating on carbonaceous surfaces can very well be approximated as spherical caps. On the other hand, we have observed a strong anisotropy
in pre-critical ice nuclei forming on the clay mineral kaolinite~\cite{Sosso_JPCL_2016_FFS-kaolinite,Sosso_ARXIV_2016_kaoliniteseed}, albeit post-critical nuclei
tended to recover the spherical cap shape. It would thus seem reasonable to build ice seeds according to the prediction of heterogeneous CNT, although nothing
prevents the user from including more exotic shapes as starting points of the RSS algorithm the HSEED methodology relies upon.

\subsubsection{Ice polytype and surface}

According to the templating effect of a particular substrate, the heterogeneous formation of ice can proceed via I$_\mathrm{c}$ or I$_\mathrm{h}$, and 
evidence of I$_\mathrm{sd}$ within the early stages of the nucleation process has also been reported~\cite{Sosso_InPreparation_2017_CHLINA}.
Moreover, for any given polytype of ice, the particular crystalline surface with which the seed interacts with the substrate has to be chosen.
Thermodynamics tells us that it is unlikely to observe high-energy (high-Miller-index) surfaces of ice
forming on any crystalline substrate. Based on a comprehensive set of previous results~\cite{Fitzner_JACS_2015_HeteroNuc-LJ}, we argue that the
following surfaces are the most plausible candidates: the basal (001),
primary prism (100) and secondary prism (110) of I$_\mathrm{h}$ and the (001) and (111) surfaces of I$_\mathrm{c}$. 
These five options represent the starting point of our RSS algorithm. Note that one could consider including additional structures in the case
of e.g. rough crystalline surfaces or defects possibly promoting the nucleation of high-Miller-index ice surfaces.
It is also worth noticing that I$_\mathrm{h}$(001) and I$_\mathrm{c}$(111) seeds expose the very same (hexagonal) plane
to the substrate, so that we expect the two seeds to give very similar results. However, we included them both in order
to assess the impact of the structural differences between I$_\mathrm{h}$(001) and I$_\mathrm{c}$(111) which emerge
within a few layers from the substrate-seed interfaces -- and consist in the different stacking of said hexagonal planes
(ABC for I$_\mathrm{c}$(111) and ABAB for I$_\mathrm{h}$(001)~\cite{Sosso_ChemRev_2016_IceReview}). 

\subsubsection{The ice-crystal interface}

More often than not, the structure of the ice nuclei at the interface with a particular crystalline substrate has very little in
common with the topology of the ice bulk phase.
For instance, density functional theory calculations have shown that the layer of water molecules mediating the
interaction between ice nuclei and the (001) surface of the mineral feldspar 
does not resemble an ice-like structure~\cite{Pedevilla_JPhysChemC_2016_feldspar-DFT}. Similar results
were obtained by means of classical force fields~\cite{Cox_PCCP_2012_lattice-match} and coarse-grained
potentials~\cite{Fitzner_JACS_2015_HeteroNuc-LJ, Bi_JPhysChemC_2016_CrystallinityHydrophilicity} as well. 
Pinpointing the structure of the one, or the more than one layers of water in contact with both the ice seed and the 
substrate is perhaps the most challenging task one has to tackle in order to extend the scope of seeded MD 
to heterogeneous ice nucleation. This is especially true when specific functional groups of the substrate (such as hydroxyl groups)
offer the possibility for supercooled water to form a hydrogen bond network between the substrate and ice. In this scenario, which is 
often observed for water in contact with a variety of potent ice nucleating agents, we have to screen as many
configurations of said hydrogen bond network as possible. 

An alternative route consists of utilising the results of enhanced sampling simulations.
For instance, in Ref.~\citenum{Sosso_ARXIV_2016_kaoliniteseed} we used metadynamics simulations to generate I$_\mathrm{h}$ and $I_\mathrm{c}$ seeds in contact with
a specific crystalline surface of the clay mineral kaolinite, and the FFS simulations of Ref.~\citenum{Sosso_InPreparation_2017_CHLINA}
provided the structure of ice seeds on CHL crystals. One could thus in principle use the preliminary results of these computationally expensive methods (e.g. non-converged
metadynamics runs or the initial interfaces only of the forward flux algorithm) as the starting point for seeded MD simulations, but this approach turns
out to require an awful lot of computational power nonetheless.

\subsection{\label{sec:HSEED_D} The HSEED method}

The HSEED methodology takes advantage instead of the RSS algorithm described in 
Refs.~\citenum{Pedevilla_JPhysChemC_2016_feldspar-DFT} and~\citenum{Kiselev_Science_2016_Fsp100Nuc}. A schematic of the HSEED work flow is
shown in Fig.~\ref{FIG_1}. We have made available via a public GitHub
repository~\cite{g_repo} a collection of (Python) scripts that can be used to apply the HSEED method to an arbitrary crystalline substrate.

\noindent\textbf{Step 1}: Spherical caps of either I$_\mathrm{h}$ or I$_\mathrm{c}$ (I$_\mathrm{sd}$ seeds can also
be considered if large enough to allow for the stacking disorder to be properly represented) are built, exposing a specific low-Miller-index surface of
the ice crystal (see above) to the substrate and containing a given number of water molecules.
These seeds are constructed directly from bulk-ice structures fulfilling the ice rules.
Seeds of different size can be built to study ice nucleation at different temperatures.
As a rule of thumb, in absence of any reference the initial size of the seeds could be
chosen as $\frac{N_\mathrm{\mathrm{C,homo}}^{*}}{2}$, i.e. half the number of water molecules contained in the homogeneous
critical nucleus size at the temperature of interest; this would be the size of an ideal heterogeneous seed displaying
a contact angle of $\sim$ 90\degree with respect to the substrate.

\noindent\textbf{Step 2}: The location of the seeds $\bm{r_{\text{Seed}}}$ as well as their relative orientation
$\phi_{\text{Seed}}$ with respect to the surface of the substrate are sampled. This step is important, as specific
structural features of the substrate can favour particular orientations of the ice
crystals~\cite{Kiselev_Science_2016_Fsp100Nuc,Sosso_ARXIV_2016_kaoliniteseed}. Then, for every
$\{\bm{r_{\text{Seed}}},\phi_{\text{Seed}}\}$ combination, we generate via the RSS procedure described in
Refs.~\citenum{Pedevilla_JPhysChemC_2016_feldspar-DFT} and~\citenum{Kiselev_Science_2016_Fsp100Nuc} a substantial number
(of the order of 10$^3$-10$^4$) of random configurations, varying the position and orientation of each water molecule within a
certain distance ($d_{\text{HB}}$ in Fig.~\ref{FIG_1}) from the surface. This procedure allows to explore the
configurational space of the hydrogen bond network between the ice seeds and the substrate. The portion of the seed
involved in the RSS typically extends up to the position of the first minimum of the density profile of water in contact
with the substrate. Subsequently, the structure of the first few layers of water in contact with the surface is
optimised via inexpensive algorithms such as the l-BFGS~\cite{d70f72bb6d4749a5a8e29137299f32ed}, keeping both the upper
part of the spherical ice cap and the substrate ``frozen''. This is because of the large surface area of the
seed-vacuum interface, which would lead to a substantial relaxation of the whole seed.
Then, we select the few structures to be used as the starting point for the seeded MD runs adopting
two criteria: (i) the topology of the seed should fit the structure of the surface as much as possible - i.e.
the number of close contacts between seed and surface should be kept at a minimum; and (ii) the structure of the seed
should be as energetically stable as possible.

\noindent\textbf{Step 3.} The selected configurations (seed plus substrate) are immersed in water, and a protocol
similar to the one used in the homogeneous case is used~\cite{Sanz_JACS_2013_MD-homonuc,
Espinosa_JChemPhys_2016_Homo-Seeding} to performed seeded MD (Step 4. in Fig.~\ref{FIG_1}). This framework involves a
cooling ramp, followed by an additional equilibration. Note that the
entire seed is kept frozen during these preliminary MD runs, in order to equilibrate the ice/water and substrate/water
interfaces without disrupting the seed-substrate interface -- which we have in any case optimised beforehand. At this
point, the HSEED methodology has brought us to a situation identical to that of the homogeneous case: we are in
possession of a few different ice seeds in contact with the substrate, and the time evolution of the system will be
monitored by means of standard MD runs at different temperatures in order to pinpoint the critical nucleus size.

\vspace{0.5cm}

Importantly, the HSEED approach allows one to rapidly obtain information about the stability of different ice faces on a given substrate.
This is crucial to heterogeneous ice formation, as being able to identify the active sites that nucleate ice on a given
substrate is perhaps the most pressing challenge in the field.
In fact, these active sites rarely seem to coincide with the low energy surfaces of
crystalline substrates. On the mineral feldspar, for example, the active sites were recently suggested to be the high energy (100)
surfaces~\cite{Kiselev_Science_2016_Fsp100Nuc}. This surface will not be exposed macroscopically on a feldspar crystal,
but will only be found within nanometric defects such as crystalline cracks and edges. 
If one wants to understand the ice nucleating efficiency of any material at a microscopic level, being able to identify where on the
surface which type of ice grows is arguably the most important piece of the puzzle. 

In addition, a qualitative estimate of the heterogeneous critical nucleus size can be made. However, we have recently
shown~\cite{fitzner2017pre} that CNT must be extended to take into account the heterogeneous nucleation of crystalline
polytypes different from the outcome of homogeneous freezing. As such, accurate references in terms of the homogeneous
critical nucleus size at different temperatures and for different polytypes are in principle needed, thus limiting the
quantitative capabilities of the HSEED method. On the other hand, this technique represents a fast route toward the
characterisation of the ice nucleating ability of whole libraries of crystalline compounds. 

In the following section we will consider the nucleation of ice on LJ as well as CHL crystals. The former represent model substrates that allow one
to extract general insight into the nucleation process, while the latter are active ice nucleating agents which have been the focus of
recent experimental work~\cite{Sosso_InPreparation_2017_CHLINA}.

\subsection{\label{sec:MD} Molecular Dynamics: Computational Details} 

In this section we describe the computational setup and the simulations performed on each class of substrate.

\subsubsection{\label{sec:MD_MW} mW water on Lennard-Jones crystals}

\begin{figure}[htbp]
\centering
\includegraphics[width=8.5cm]{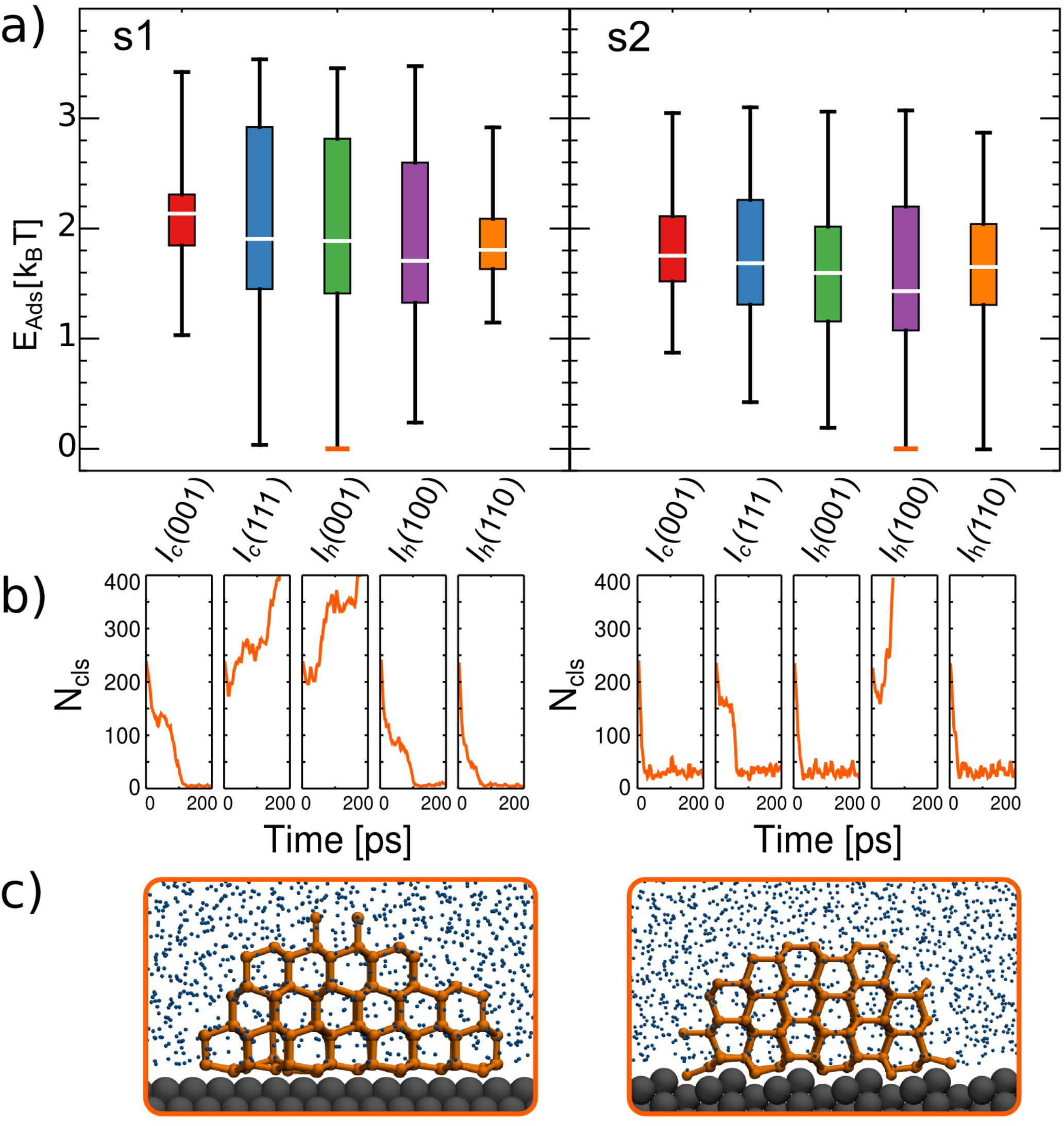}
\caption{a) Adsorption energy per water molecule in the contact layer of different ice seeds ($\sim$ 250 molecules per
seed) on the two substrates used with the mW model. The lower (upper) end of the whisker boxes and
the white line within stand for the 25$^\mathrm{th}$(75$^\mathrm{th}$) percentile and the median of the data, respectively. The
lower (upper) end of the error bars corresponds instead to the energy of the most (least) stable
structure. b) Number of water molecules in the ice seeds $N_{\text{cls}}$ as a function of time, as 
obtained in seeded MD simulations of the most stable seed found via RSS for each of the ice polytype/faces combinations 
illustrated in panel a). c)Representative snapshot of the most stable ice seeds on s1 (left panel) and s2 (right panel). Substrate, ice
seed, and liquid water are depicted in gray, orange and blue, respectively.}
\label{FIG_2}
\end{figure}

We considered in the first instance the heterogeneous freezing of the coarse-grained mW model for
water~\cite{Molinero_JPCB_2008_mW}. In this case, water is represented by a single bead (there are no explicit hydrogen
atoms) and interacts with other water molecules via a three-body potential that favours tetrahedral order. We have taken
advantage of this water model in previous studies aimed at understanding the ice nucleation capabilities of
idealised~\cite{cox2015molecular1,cox2015molecular2,fitzner_many_2015,fitzner2017pre} and hydroxylated model
surfaces~\cite{pedevilla2017whatmakes}. In order to validate our seeding approach we have chosen two particular fcc
surfaces (labelled s1 and s2) which interact with the water via a Lennard-Jones potential (details can be found
in Ref.~\citenum{fitzner2017pre}). In our previous work we employed metadynamics
simulations~\cite{laio2002escaping,barducci2008well} to establish what sort of ice nuclei form on the s1 and s2 surfaces
at a temperature of 235 K. As we took advantage of a collective variable (PIV~\cite{gallet2013structural}) which is free
from bias toward any particular ice polytype or crystalline face, we have unequivocally determined that s1 and s2
promote the heterogeneous nucleation of I$_\mathrm{h}$(001)/I$_\mathrm{c}$(111) and I$_\mathrm{h}$(100), respectively. We have also obtained an
estimate of the critical nucleus size: $211 \pm 11$ and $104 \pm 3$ water molecules for s1 and s2, respectively.   

By comparing the results of Ref.~\citenum{fitzner2017pre} to the outcomes of the HSEED approach we will thus have the
opportunity to validate both the predictive power and the accuracy of the HSEED methodology. Moreover, the mW/LJ
computational setup is much less expensive compared to the simulations of ice formation on CHL crystals (see next
Section). We thus have the possibility to assess the impact on the HSEED method of intrinsic variables such as the size
of the seeds and temperature. To this end we start by performing a RSS for the five combinations of ice polytype/faces
considered in this study (see Section~\ref{sec:HSEED}) in contact with either s1 and s2, varying the number of molecules
in the seeds from 50 to 400 (in increments of 50). From the resulting dataset upon energy minimisation, we select
three seeds according to the two criteria specified in Section~\ref{sec:HSEED}, solvate the latter in a slab of water
(so as to reach $\sim$ 4000 water molecules in the whole of the simulation box) and proceed to perform twenty seeded
MD runs for each seed. The production runs followed a 0.2 ns long equilibration of the systems at 273 K, where the
molecules within the seeds are kept frozen, and a subsequent quenching to the target temperature within 2 ns. We
sampled the NVT ensemble by means of a ten-fold Nos\'{e}-Hoover chain~\cite{martyna_nosehoover_1992} with a relaxation
time of 1~ps and a timestep of 10~fs using the LAMMPS package~\cite{plimpton_fast_1995}. As opposed to the fully
atomistic water models, when dealing with mW water the outcome of the seeding runs can almost be considered as binary,
in that we observe either the very rapid freezing of the whole water slab within a few nanoseconds, or the complete
dissolution of the seed within short timescales. We shall see in Section~\ref{sec:TIP4P} that in order to observe the
growth of ice nuclei on CHL crystals we will need instead to monitor the seeds for as long as hundreds of nanoseconds.

\subsubsection{\label{sec:MD_CHL} TIP4P/Ice water on cholesterol crystals}

We also applied the HSEED approach to investigate ice nucleation on cholesterol
monohydrate~\cite{Craven_Nature_1976_CHLMonohydrate} (CHLM). 

A single layer of CHL molecules, cleaved along the (001) plane (perpendicular to the normal to the slab) 
was prepared by starting from the experimental cell parameters and lattice positions~\cite{Craven_Nature_1976_CHLMonohydrate}.
Specifically, a CHLM crystal system made of two mirroring slabs (intercalated by
water molecules, in a ratio of 1:1) was cleaved along
the (001) plane. The triclinc symmetry of the system (space group $C1$) was preserved, and we have constructed
a 3 by 3 supercell with in-plane dimensions of 37.17 and 36.57 \AA.
We positioned 1923 water molecules randomly atop this CHLM slab at the density of the TIP4P/Ice
model~\cite{Abascal_JCP_2005_TIP4P-ice} at 300 K, and expanded the dimension of the simulation cell along the normal to the
slab to 100 \AA.

10$^3$ structures for I$_\mathrm{h}$(001), I$_\mathrm{h}$(100), I$_\mathrm{h}$(110), I$_\mathrm{c}$
(001) and I$_\mathrm{c}$(111) seeds were generated, each one containing $\sim$ 250 water molecules. The energy minimisations
were performed via the GROMACS MD package~\cite{Hess_JCTC_2008_GROMACS-4, VDSpoel_JCC_2005_GROMACS-3} using the
CHARMM36~\cite{Bjelkmar_JChemTheoryComput_2010_CHARMM, Lim_JPCB_2011_CHARMMCHLUpdate} and
TIP4P/Ice~\cite{Abascal_JCP_2005_TIP4P-ice} force fields to describe CHLM and water molecules respectively. A
validation of this particular setup can be found in Ref.~\citenum{Sosso_InPreparation_2017_CHLINA}. According to the
criteria illustrated in section~\ref{sec:HSEED}, three seeds for each ice polytype/surface (e.g. I$_\mathrm{h}$(001)) were
selected following the outcome of the RSS procedure. These seeds have been immersed in a $\sim$ 45 \AA$\ $ thick water slab, 
which resulted in simulation boxes containing $\sim$ 2000 water molecules.

MD simulations have also been performed using the GROMACS package. The equations of motion were integrated via a leap-frog algorithm, with a
timestep of 2 fs. Electrostatic interactions were treated by means of a particle-mesh Ewald
summation~\cite{Essmann_JCP_1995_PME} with a cutoff of 12 \AA. Non bonded interactions were calculated up to 10 \AA,
and a switching function was used to bring them to zero at 12 \AA. We sampled the NVT ensemble using a stochastic velocity
rescaling thermostat~\cite{Bussi_JCP_2007_v-rescale} with a coupling constant of 2 ps. The rigid geometry of TIP4P/Ice
molecules was enforced thanks to the SETTLE algorithm~\cite{Miyamoto_JComputChem_1992_SETTLE}, while additional
constraints were treated via the P-LINCS
algorithm~\cite{Hess_JComputChem_1997_LINCS, Hess_JCTC_2008_PLINCS}.

The equilibration of the substrate/water and ice/water interface started with a 5 ns run at 300 K, followed by a 5
ns-long cooling ramp from 300 to 200 K. A 2 ns long equilibration at 200 K followed, after
which seeded MD production runs were performed at the desired target temperature by
randomly selecting the initial atomic velocities according to the corresponding Maxwell-Boltzmann distribution.
To verify the consistency of our results, we run simulations involving multiple seeds, different initial velocities, and 
different cell sizes. We shall see that the HSEED method provides a robust set of results.

\section{\label{sec:RESULTS}Results}

\subsection{\label{sec:MW} mW water on Lennard-Jones crystals}

We start by focusing on the case of mW water freezing on the LJ crystals s1 and s2 described in Section~\ref{sec:MD_MW}.

{\bf Step 1} We built ice seeds of different sizes (containing from 50 to 400 in increments of 50 water molecules) choosing five
combinations of crystal polytype and face exposed to the substrate: I$_\mathrm{h}$(001), I$_\mathrm{h}$(100), I$_\mathrm{h}$(110), I$_\mathrm{c}$(001)
and I$_\mathrm{c}$(111).

\begin{figure}
\centering
\includegraphics[width=8cm]{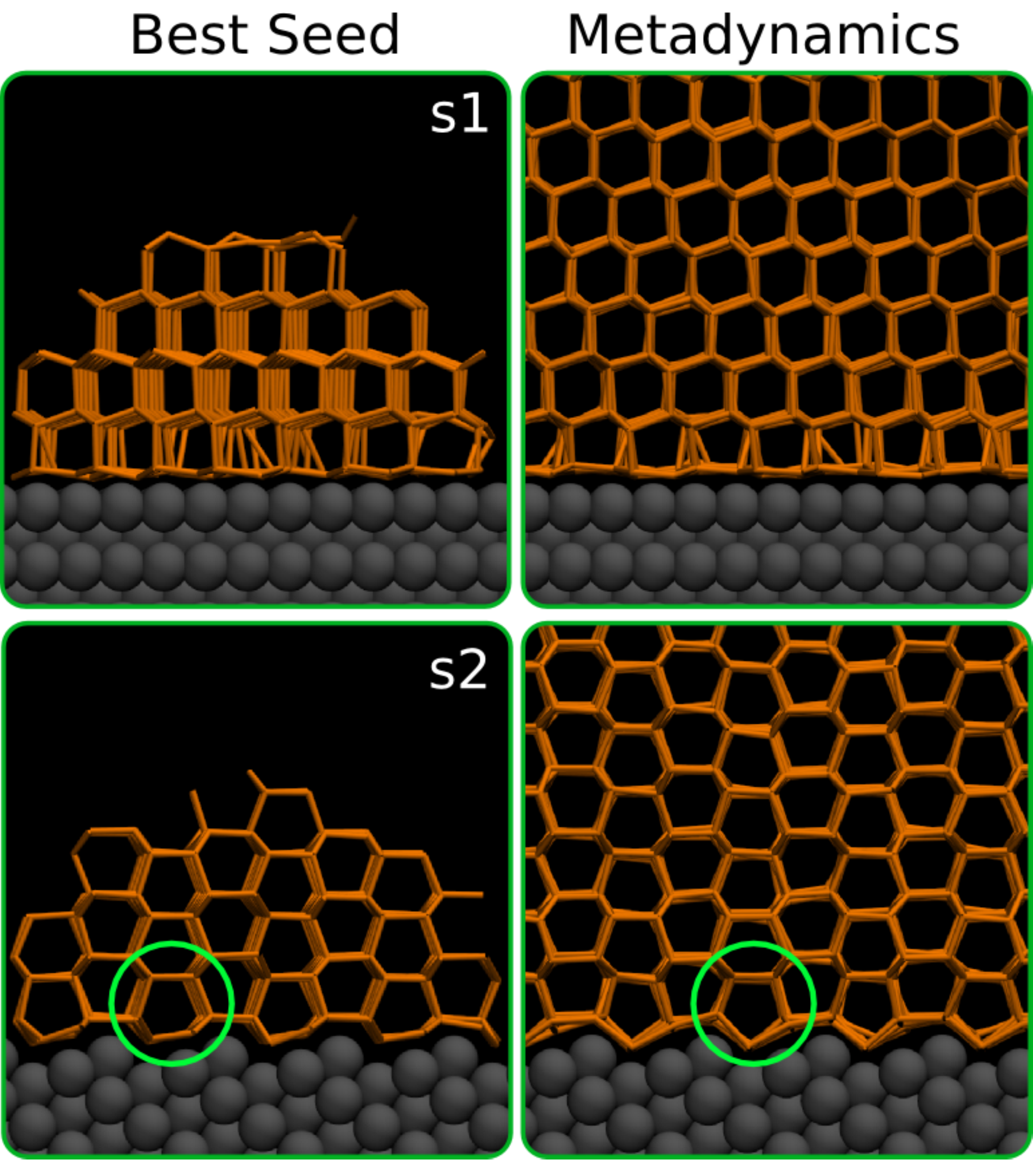}
\caption{The most stable seed selected via the RSS algorithm for s1 (top left) and s2 (bottom left), compared with the
outcome of metadynamics simulations~\cite{fitzner2017pre} (top right and bottom right for s1 and s2, respectively).
Bonds between water molecules within the ice nuclei and s1/s2 atoms are shown in orange and grey, respectively. The green circles
highlight the small difference between the two approaches in terms of the structure of the contact layer of the seeds on s2.
Note that the orientation of the best seed in both cases is the same as the one found in metadynamics.}
\label{FIG_3}
\end{figure}

{\bf Step 2} We generated by means of our RSS algorithm between 5,000 and 30,000 ice seeds for each combination of seed
size, ice polytype and ice crystalline face (see Fig.~\ref{FIG_1}), exploring different locations and orientations
of the ice seed on the substrate as well as optimising the geometry of the seed-substrate interface. The adsorption
energy per water molecule $\text{E}_{\text{Ads}}$ in the contact layer (i.e. within 4~\r{A} of the substrate) for each
type of ice seed (in this case containing $\sim$ 250 water molecules) as obtained upon energy minimisation is shown in
Fig.~\ref{FIG_2}. As expected, the spread of $\text{E}_{\text{Ads}}$ is huge. 
We remark that this spread should not be considered as a source of uncertainty: on the contrary, it 
represents a measure of the extent to which the configurational space for a given seed as been explored. As
such, a large spread is actually desirable, and the evolution of it as the RSS progresses provides an
indication of the convergence of the algorithm. We also note that the lowest value of $\text{E}_{\text{Ads}}$ 
found by the RSS for a given seed is the quantity that matters in determing the relative stability of different seeds - which
in this respect can differ by as much as 1 $k_BT$ (see e.g. I$_\mathrm{c}$(001) and I$_\mathrm{c}$(111)) in Fig.~\ref{FIG_2}a.
\sss{However, the} As illustrated in Fig.~\ref{FIG_2}a, I$_\mathrm{h}$(001)/I$_\mathrm{c}$(111) and
I$_\mathrm{h}$(100) seeds are amongst the most stable ones for s1 and s2, respectively. As illustrated in Fig.~\ref{FIG_3}, these
seeds correspond to the outcome of previous metadynamics simulations (see Section~\ref{sec:MD_MW}). The morphology of
the seeds, for instance in terms of $\bm{r_{\text{Seed}}}$ and $\phi_{\text{Seed}}$ (see Section~\ref{sec:HSEED}), is
correctly reproduced by the HSEED framework (see Fig.~\ref{FIG_3}). A small mismatch between metadynamics and HSEED can
be observed for the contact layer of the seeds on s2: although the network of water molecules is aligned correctly, the
contact layer in the trenches does not exactly match the one obtained via metadynamics. We will make use of
this observation to evaluate the overall robustness of the HSEED approach later on.

\begin{figure*}
\centering
\includegraphics[width=16cm]{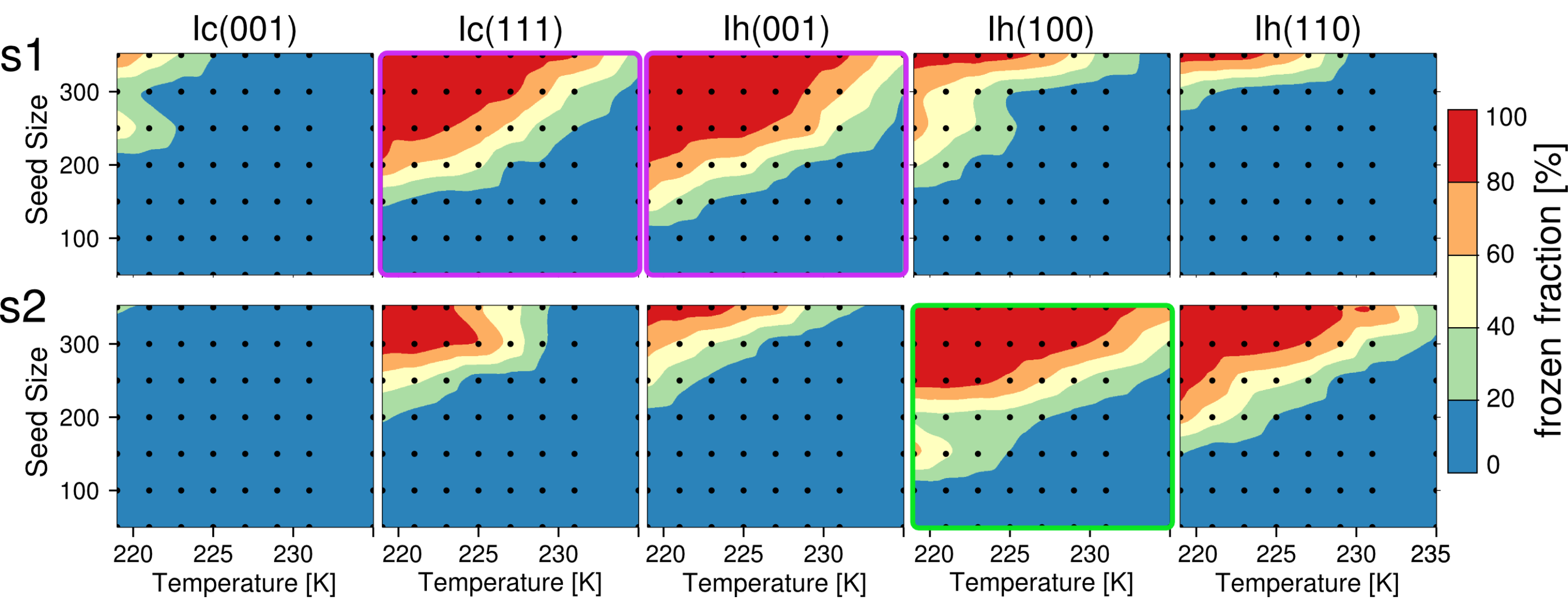}
\caption{Frozen percentage of simulations as a function of the temperature and seed size. Each black point indicates
a set of twenty seeded MD runs at a given temperature, starting from the three best structures for a given seed size.
The colour map represents the frozen percentage - i.e. the percentage of simulations where the ice seed grew to fill the
whole simulation box as opposed to dissolve - for each collection of seeded MD runs. To generate 
smooth two dimensional maps we applied cubic interpolation between data points. The purple and green frames highlight the combinations
of ice polytype/face we have observed nucleating on s1 and s2, respectively -- by means of metadynamics
simulations~\protect\cite{fitzner2017pre}.}
\label{FIG_4}
\end{figure*}

In light of the outcomes of the RSS algorithm, one could be tempted to draw the conclusion that the most stable types of
seeds (e.g. I$_\mathrm{h}$(001) and I$_\mathrm{c}$(111) for s1 in Fig.~\ref{FIG_2}), as obtained upon energy minimisation, would have the
highest probability to grow on a given substrate. However, we shall see in Section~\ref{sec:TIP4P} that this is not
always the case. In fact, in order to assess which particular ice polytype and face would be favoured the most on a
specific substrate, we have to use the seeds as the starting point for seeding MD simulations.

{\bf Step 3} We picked the three ``best'' structures from the RSS dataset (according to the criteria specified in
Section~\ref{sec:HSEED}) for each ice polytype/face and seed size, solvated them and performed twenty MD runs at
different target temperatures (see Section~\ref{sec:MD_MW} for further details). The results are summarised in
Fig.~\ref{FIG_4}: it is clear that for a low enough temperature and reasonable seed size most of the polytype/face
combinations will initiate freezing within a substantial fraction of the MD runs. At higher temperatures, however, only
the ''correct`` (i.e. the same observed via the metadynamics simulations of Ref.~\citenum{fitzner2017pre}) crystal face
is capable of promoting the formation of ice.  We note that in the case of s2 the secondary prism face of I$_\mathrm{h}$ is also
a reasonable candidate - in agreement with the findings of our previous work~\cite{fitzner_many_2015}. Importantly, the
above mentioned small metadynamics-HSEED mismatch in terms of the contact layer for I$_\mathrm{h}$(100) seeds on s2 does not seem
to impact the outcomes of the HSEED method.

As shown in Fig.~\ref{FIG_4}, for s2 the 400-molecule seed seems to be less effective in promoting ice formation than a
350-molecule seed. This is due to an artefact of the RSS as we have visually verified that the 400-molecule
seeds have the ``wrong'' orientation on the s2 surface if compared to the 350-molecule ones. This is because we have not generated
enough 400-molecule seeds to properly sample the configurational space - due to the computational cost of the RSS for
large seeds. Such artefacts can be avoided by parallelising the (to date serial) RSS algorithm and introducing
additional criteria for the selection of the seeds, possibly based on order/disorder parameters.

We have also found that only seeds that are substantially larger than the critical nucleus size estimates obtained in
Ref.~\citenum{fitzner2017pre} induce nucleation on both s1 and s2. Specifically, according to the HSEED method 
$N_\mathrm{C}^{*}$ at 235 K is equal to 330$\pm$25 and 290$\pm$ for s1 and s2, respectively -- to be compared with 211$\pm$11
and 104$\pm$3 for s1 and s2 respectively, as obtained in Ref.~\citenum{fitzner2017pre}.
This is most likely to do with: (i) the structure of
the ice seed; a crystalline surface interface as obtained via the HSEED method, even upon minimisation, is bound to be
more defective than that obtained via conventional enhanced sampling techniques (metadynamics included); (ii) the short
re-equilibration of the water-seed interface (see Section~\ref{sec:MD_MW}) negatively impacts the freezing probability
of the seed; (iii) the assumption of a contact angle that is likely to be larger than that of the nuclei 
obtained via e.g. metadynamics simulations.
In the case of s2, where the discrepancy in terms of $N_\mathrm{C}^{*}$ between HSEED and metadynamics amounts to almost
a factor two, we have found that indeed the critical nuclei obtained via metadynamics are on average rather flat and 
characterised by small contact angles (of the order of $\sim$45\degree). Exploring different contact angles as an
additional degree of freedom within the HSEED method will be the subject of future work. However, we note that the 
relative trends in terms of the critical nucleus size are consistent in that $N_\mathrm{C}^{*} (s1) > N_\mathrm{C}^{*} (s2)$
according to both HSEED and metadynamics. Moreover, our results suggest that screening different contact angles is not 
necessary to establish which polytype/face will form on a particular substrate. 
Finally, we remark that, in the case of mW water, longer equilibration times for the seeds are difficult to deal with,
because the fast dynamics of the model is likely to induce heterogeneous freezing within relatively short time scales --
notwithstanding the particular morphology of the seed. 

Further evidence of the net preference for the s1 and s2
surfaces to promote the formation of I$_\mathrm{c}$(111)/I$_\mathrm{h}$(001) and I$_\mathrm{h}$(100) is provided by the distribution of the
potential energies of each one of the seeded MD runs. Specifically, we find that the systems seeded with the
``correct'' crystal face (see FIG.~\ref{FIG_3}) are characterised on average by the lowest potential energy 
after freezing of all the water molecules in the simulation cell.
This suggests that the I$_\mathrm{c}$(111)/I$_\mathrm{h}$(001) and I$_\mathrm{h}$(100) seeds in the case of
s1 and s2 respectively led to the formation of more pristine ice if compared to the other polytype/face
combinations.

For the purpose of establishing these trends we accumulated a total of 76.8 $\mu$s of simulation time (two systems
$\times$ five ice faces $\times$ eight seed sizes $\times$ eight temperatures $\times$ three seeds $\times$ twenty MD
runs $\times$ 2~ns simulation time). However, if one would be interested in (i) pinpointing the most probable seed
morphology; and (ii) obtaining an estimate of the critical nucleus size for a given substrate at a given temperature,
only a small fraction of this computational effort would be needed.  Our results suggest that in this case one would
need about 0.5 $\mu$s.

\subsection{\label{sec:TIP4P} TIP4P/Ice water on cholesterol crystals}

The freezing of mW water on the LJ crystals just discussed allowed us to explore the capabilities of the HSEED method for
a variety of nucleation scenarios/conditions. However, the true testing ground is heterogeneous nucleation of ice from fully atomistic water
models on complex/realistic crystalline surfaces, a situation where enhanced sampling simulations are necessary to observe even a single nucleation event - often
requiring phenomenal computational resources. As such, we have applied the HSEED method to the formation of ice on
CHLM crystals; a problem which we have 
recently tackled with (computationally expensive) FFS simulations~\cite{Sosso_InPreparation_2017_CHLINA}. Specifically, we consider
the (001) hydroxylated surface of CHLM crystals (CHLM$^{-OH}_{001}$), as detailed in Sec.~\ref{sec:MD_CHL}.

\begin{figure}
\begin{centering}
\includegraphics[width=8cm]{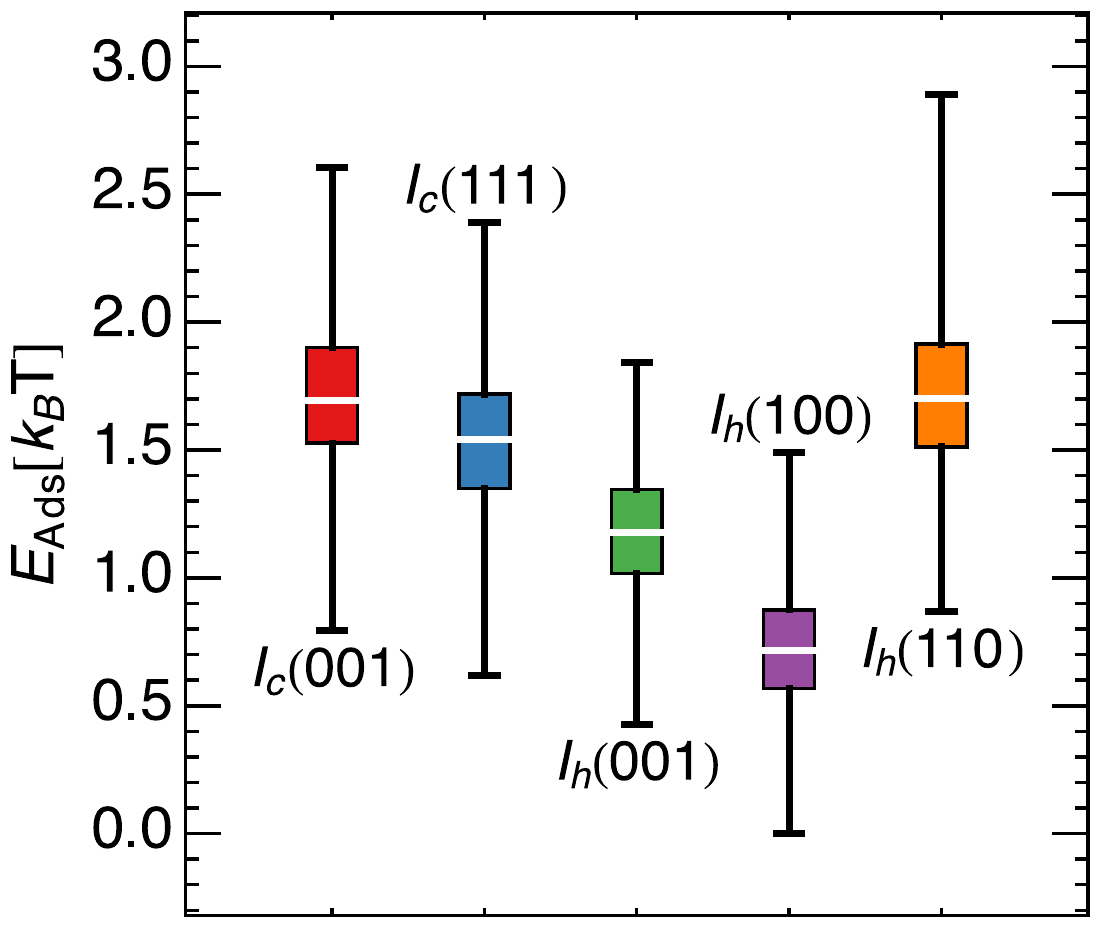}%{energetics_after_minimization.png}}
\par\end{centering}
\caption{Adsorption energy per water molecule in the contact layer of different ice seeds ($\sim$ 250 molecules per seed) on CHLM$_{\text{001}^{-OH}}$.
The lower (upper) end of the whisker boxes and the white line within stand for the 25$^\mathrm{th}$(75$^\mathrm{th}$) percentile and the median of the data, respectively. The lower (upper) end of the error bars corresponds instead to the energy of the most (least) stable structure.}
\label{FIG_5}
\end{figure}

{\bf Step 1} The same five combinations of ice polytype/face detailed in the previous section have been considered as the starting point for the HSEED procedure.
Guided by the outcome of our FFS simulations~\cite{Sosso_InPreparation_2017_CHLINA}, we built seeds containing 250 water molecules -- roughly the
dimension of ${N_\mathrm{\mathrm{C,hetero}}^{*}}$ at 230 K.

{\bf Step 2} About 2,000 structures for each seed have been generated via the RSS algorithm detailed in Sec.~\ref{sec:HSEED}. 
The average adsorption energy per water molecule $E_{\text{Ads}}$ for the different ice polytype/face combinations
as obtained upon energy minimisation, is shown in Fig.~\ref{FIG_5}. Similar to what we observed for mW water on LJ crystals, the spread of these
data is huge. Interestingly, the most energetically stable seeds found expose the I$_\mathrm{h}$(100) and I$_\mathrm{h}$(001) surfaces at the ice-CHLM$^{-OH}_{001}$ interface, while
our FFS simulations~\cite{Sosso_InPreparation_2017_CHLINA} unequivocally pinpointed I$_\mathrm{c}$(100) nuclei as the kinetically more favoured to form
on CHLM$^{-OH}_{001}$. This is in contrast with what we have observed in the case of mW water on LJ crystals, where the most stable ice seeds displayed the
same morphology as those obtained via metadynamics simulations. Is the HSEED thus incapable of dealing with complex interfaces such as the 
ice-CHLM$^{-OH}_{001}$ one? To answer this question we kept following the work flow of the HSEED method (see Fig.~\ref{FIG_1}).

\begin{figure} 
\begin{centering}
\centerline{\includegraphics[width=8.0cm]{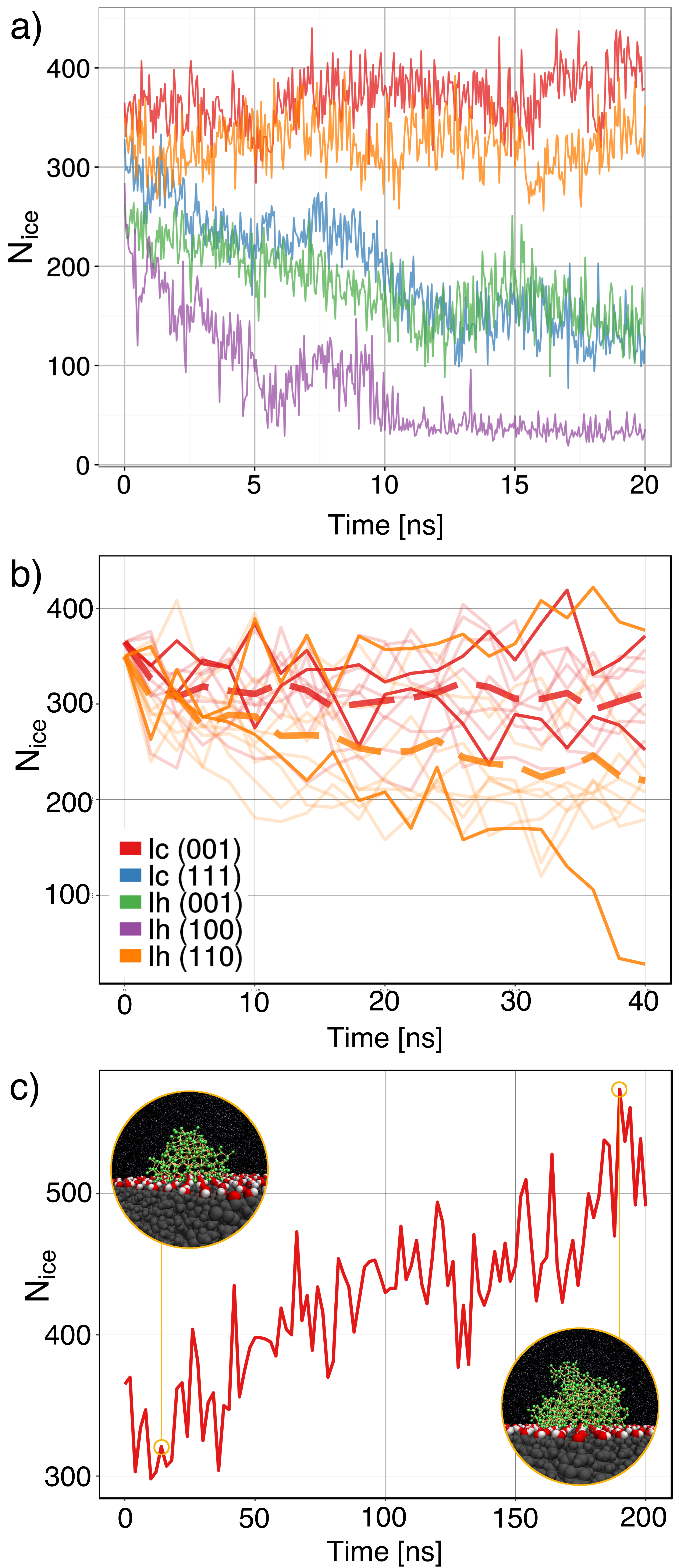}}
\par\end{centering}
\protect\caption{a) Number of molecules within different (see legend) ice seeds on the CHLM$^{-OH}_{001}$ surface as function of time.  
b) Same as panel a) for ten statistically independent simulations of I$_\mathrm{c}$(001) and I$_\mathrm{h}$(110) seeds. 
The curves corresponding to the simulations
leading to the biggest and smallest seeds are shown with thick continuous lines, while thick dashed lines correspond to the mean size
of the seed at any given point in time.  
c) Growth of an I$_\mathrm{c}$(001) seed over a longer timescale (200 ns). The insets show representative snapshots of small (left) and large (right) 
seeds.}
\label{FIG_6}
\end{figure}

{\bf Step 3} We selected three seeds for each ice polytype/face combination according to the criteria specified in Section~\ref{sec:HSEED}, and embedded them
in a slab of liquid water. The equilibration protocol preceding the seeding MD runs is described in Section~\ref{sec:MD_CHL} and led 
to a substantial increase in the size of the seeds, from 250 to $\sim$ 350 molecules.
We have chosen to perform seeded MD simulation at 240 K, as at this temperature the dynamics of liquid water is reasonably fast -- while the critical
nucleus should be of the order of 200-300 water molecules, according to our FFS simulations~\cite{Sosso_InPreparation_2017_CHLINA}.
The outcome of these simulations is summarised in 
Fig.~\ref{FIG_6}a: I$_\mathrm{c}$(111), I$_\mathrm{h}$(001) and I$_\mathrm{h}$(100) seeds dissolve within 20 ns, while I$_\mathrm{c}$(001) and
I$_\mathrm{h}$(110) seeds endure. The same trend can be
observed for different configurations of the initial seeds as well for different choices of the initial velocities. As an example, we report in 
Fig.~\ref{FIG_6}b additional sets of simulations for I$_\mathrm{c}$(001) and I$_\mathrm{h}$(110) seeds: despite an initial drop in the number of molecules within the
seeds (which is due to the sub-optimal equilibration of the seed/water and seed/CHLM interfaces), these two combinations of ice polytype and face seem to be
stable, on average, up to 40 ns. Note that, as opposed to the mW water on the LJ crystals, the time
scales involved for the growth and dissolution of
the seeds are much longer. Nonetheless, we were able to probe the actual growth of the stable ice seeds employing only a
fraction of the computational
effort of the FFS simulations of Ref.~\citenum{Sosso_InPreparation_2017_CHLINA}. We found that, consistently with the latter,
I$_c$(001) seeds do grow, as illustrated in Fig.~\ref{FIG_6}c. In addition, the HSEED result in terms of the critical nucleus size
($N_\mathrm{C}^{*}$=350$\pm$50 at 240 K) is compatible with the outcome of our FFS simulations ($N_\mathrm{C}^{*}$=250$\pm$50 at 230 K).

These results indicate that the RSS alone is not sufficient to determine which ice
polytype and face would be favoured on a specific substrate. 
Such insight has to be gained from seeded MD
simulations, thus illustrating the importance of each step in the HSEED framework. In addition,
the values of $E_{\text{Ads}}$ reported in Figs.~\ref{FIG_2},~\ref{FIG_5}
originate not only from the interaction between the ice seeds and the crystalline substrate, but also from the surface energies of
the different ice crystalline faces. For instance, the two low energy surfaces of hexagonal ice (I$_\mathrm{h}$(001) and I$_\mathrm{h}$(100)) are more
stable than the secondary prism face, I$_\mathrm{h}$(110), of hexagonal ice.

It is also intriguing to note that, while I$_\mathrm{c}$(001) seeds are the most kinetically favourable at this strong supercooling,
I$_\mathrm{h}$(110) nuclei are also possible. This is consistent with the results of Ref.~\citenum{Sosso_InPreparation_2017_CHLINA}, which have
shown that CHLM crystals can promote the formation of both I$_\mathrm{c}$ and I$_\mathrm{h}$ pre-critical nuclei. In fact, our FFS 
simulations~\cite{Sosso_InPreparation_2017_CHLINA} suggest that a coexistence of the two polytypes can be expected at mild supercooling.
The HSEED method thus provides further support to this hypothesis, which is in stark contrast to what has been observed in terms of ice formation
on several inorganic crystals. For instance, according to both experiments and simulations, exclusively I$_\mathrm{h}$(100)
forms on both the clay mineral kaolinite~\cite{Cox_FaradayDiscuss_2013_MD-kao-nuc,
Zielke_JPCB_2015_kaolinite-IN, Sosso_JPCL_2016_FFS-kaolinite} and the mineral feldspar~\cite{Kiselev_Science_2016_Fsp100Nuc}. 

\begin{figure} 
\begin{centering}
\centerline{\includegraphics[width=8.0cm]{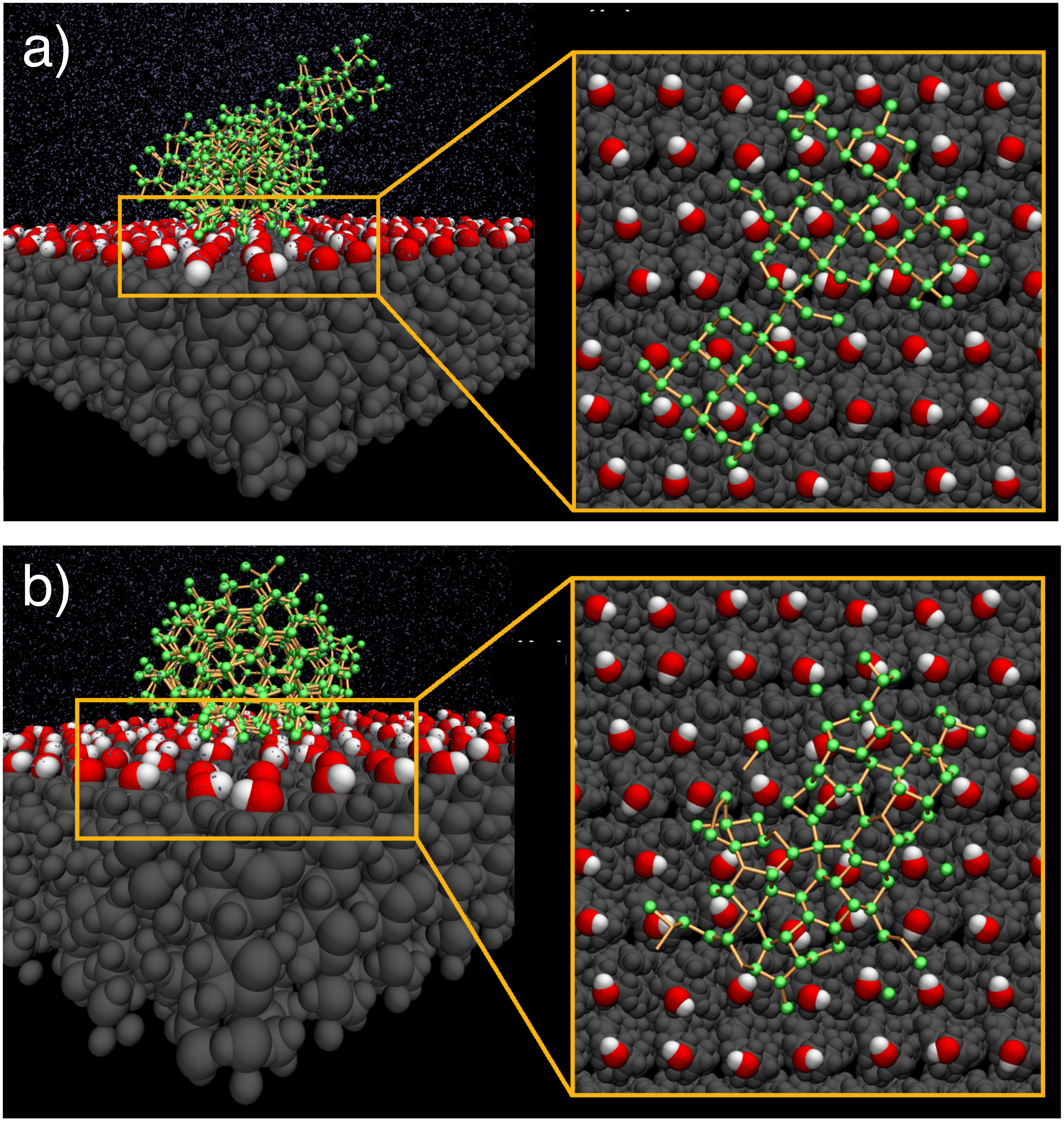}}
\par\end{centering}
\protect\caption{Representative snapshot of a) a I$_\mathrm{c}$(001) seed and b) a I$_\mathrm{h}$(110) seed at 230 K
(side/top view on the left/right), growing on the CHLM$^{-OH}_{001}$ surface during seeded MD simulations. Water
molecules not participating in the ice nuclei are not shown. CHL molecules, the oxygen(hydrogen) atoms of their hydroxyl groups, and the
oxygen atoms of ice-like molecules are depicted in grey, red(white) and green, respectively.}
\label{FIG_7}
\end{figure}

The rare ability of CHLM crystals to accommodate both I$_\mathrm{c}$(001) and I$_\mathrm{h}$(110) seeds could be due to
the particular arrangement of the hydroxyl groups of CHL at the ice-CHLM$^{-OH}_{001}$ interface. As illustrated in Fig.~\ref{FIG_7}, this
seems indeed to be the case, as both I$_\mathrm{c}$(001) and I$_\mathrm{h}$(110) seeds tend to align along preferential directions
leading to the relevant ice faces to grow along rows of hydroxyl groups. However, water molecules at the ice-CHLM$^{-OH}_{001}$ interface 
are much more ordered for I$_\mathrm{c}$(001) seeds if compared to the I$_\mathrm{h}$(110) case. We argue that I$_\mathrm{h}$(110) seeds can be
stabilised nonetheless by the CHL surface due to the intrinsic flexibility of this substrate, which can play a significant role
in the context of the kinetics of ice formation~\cite{Sosso_ARXIV_2016_kaoliniteseed}.
For instance, the surfaces of both feldspar and kaolinite are held together by strong covalent bonds, resulting in a rather rigid surface. On
CHLM however, weak intermolecular interaction only are responsible for the stability of the surface. This is a
fundamental difference between inorganic and organic crystals, which may very well be at the heart of the strong ice nucleating ability of the 
latter~\cite{Sosso_InPreparation_2017_CHLINA,Fukuta_JPhysChemSolids_1963_INAonOrganicCrystals}.

\subsection{\label{sec:SPEEDU} Computational cost}

The challenging case of ice nucleation on CHLM represents an opportunity to compare the computational cost of the HSEED method with that
of the FFS simulations reported in Ref.~\citenum{Sosso_InPreparation_2017_CHLINA}. 
Generating 10$^3$-10$^4$ seeds for each ice polytype/face combination required $\sim$ 
48 CPU hours. The geometry optimisation of the interfacial region for each one of these seeds took - on average - 0.08 CPU hours, totalling 800 CPU hours.
Note that the minimisation runs can be trivially parallelised, so that this stage of the algorithm can typically be dealt with within a day. The bulk of the
computational effort lies within the actual seeded MD runs. Including the equilibration stage, we estimate a cost of 40 ns $\times$ 10 seeded MD runs $\times$ 3 seeds for each
ice polytype/face combination $\times$ 5 ice polytype/face combinations $\times$ 12 ns/day (using 8 CPUs) = 96,000 CPU hours. Overall, the HSEED algorithm
thus allowed us to investigate the formation of ice on CHLM at strong supercooling using $\sim$ 10$^5$ CPU hours. 

The FFS simulations reported in
Ref.~\citenum{Sosso_InPreparation_2017_CHLINA} required $\sim$ 10$^6$ CPU hours - taking advantage of GPU acceleration (providing a $\sim$ 4$\times$ speedup).
Importantly, the FFS algorithm relies on the definition of different interfaces (see e.g. Ref.~\citenum{0953-8984-21-46-463102}) 
along the path from water to ice, which have to be sampled one after the other.
The same holds to various extents for most path sampling methods. Similarly, free energy-based enhanced sampling methods
such as metadynamics can be parallelised by means of e.g. multiple walkers~\cite{doi:10.1021/jp054359r} but still rely on the sampling of the free energy surface by means
of serial production runs. On the other hand, \textit{all} the production runs within the HSEED framework can be performed in a trivially parallel fashion, so that
the computational cost of the HSEED can be dealt with much more quickly than e.g. FFS and metadynamics. 
To provide a practical example, the FFS simulations reported in Ref.~\citenum{Sosso_InPreparation_2017_CHLINA} required a year-long project, while the HSEED simulations
described here took one month only.

Interestingly, we observed a nominal speedup of about one order of magnitude in the case of mW water freezing on LJ crystals as well. In order to investigate a 
single surface at a particular temperature, the HSEED required $\sim$ 10$^4$ CPU hours, to be compared with the $\sim$ 10$^5$ CPU hours needed to converge the
metadynamics simulations of Ref.~\citenum{fitzner2017pre} for the exact same system. 
We note that, despite the substantial number of different ice seeds (in terms of size/polytype) we have probed in this case,
the RSS algorithm did not represent a limiting step: as an example, taking into account one substrate and 30 different combinations of ice seed size and polytype
only required 1 CPU for 7 (2) days when dealing with seeds containing 400 (100) molecules.
Finally, we remark that investigating ice nucleation at
mild supercooling is simply not feasible by means of conventional enhanced sampling techniques, due to the low nucleation rate. The unique strength of the HSEED thus stands in the capability of the method to address this important pitfall.

\section{\label{sec:Discussion}Discussion and Conclusions}

In summary, we have presented a methodology (HSEED) to study the heterogeneous nucleation of ice via a combination of
RSS algorithms and seeded MD simulations. We have made available via a public GitHub
repository~\cite{g_repo} a collection of (Python) scripts that can be used to apply the HSEED method to an arbitrary crystalline substrate.
We validated our approach by comparing the outcomes of the
HSEED method against enhanced sampling simulations of: (i) coarse-grained mW water freezing on model LJ crystals~\cite{fitzner2017pre}; and
(ii) fully atomistic TIP4P/Ice water turning into ice on CHLM crystals~\cite{Sosso_InPreparation_2017_CHLINA}. In both cases the HSEED method is able to
pinpoint the combination of ice polytype and crystalline face which is most likely to form on the crystalline
substrates. Estimates of the critical nucleus size are also in line with independent evaluations. 

When dealing with computationally inexpensive simulation setups such as mW water on model surfaces, the HSEED method
allows the comprehensive investigation of the ice nucleating ability of different substrates at different temperatures,
including mild supercooling for which - costly - enhanced sampling simulations would be needed. Specifically, in
this case one can think about two different approaches to look for the ``correct'' ice seed on a given substrate:

\begin{itemize}
\item \textit{Constant Seed Screening}: starting from a dataset of different ice seeds of a given size, the temperature of the whole system is lowered until 
heterogeneous ice nucleation is observed for one (or more than one) of the ice polytype/face combinations.
   \item \textit{Constant Temperature Screening}: at a given temperature, the size of different ice seeds is incrementally increased until 
heterogeneous nucleation is observed for one (or more than one) of the ice polytype/face combinations. 
\end{itemize}

The former would be the method of choice when dealing with computationally inexpensive MD runs, as only one RSS has to
be performed. The latter method might perform better if the seeding MD simulations turn out to be very expensive and/or
if it would take longer MD runs to observe nucleation events, as it would be quicker to run multiple RSSs.

Importantly, the HSEED method performed well even in the challenging case of ice formation on CHLM. In
this scenario, the hydrogen bond network between the ice seeds and the substrate had to be explicitly taken into account, and
the complexity of the ice-crystal interface provided a real testing ground for the approach. We were able to identify via the HSEED approach
the same combination of ice polytype/face we observed by means of forward flux sampling simulations~\cite{Sosso_InPreparation_2017_CHLINA}, and the structure
of the seeds-substrate interface is consistent with what we have found via brute force MD simulations~\cite{Sosso_InPreparation_2017_CHLINA}. The specific
surface of CHLM crystals we have considered in here is capable, according to previous results, to accommodate two
different ice polytypes, an evidence that the HSEED method did capture as well.

In its present formulation, this method can treat relatively flat, pristine crystalline surfaces.
This represents a substantial leap forward for the ice nucleation community, as we are now in a position to
evaluate rapidly the ice nucleation ability of whole libraries of crystalline compounds with the same
computational effort required to investigate a single substrate by means of conventional enhanced sampling methods.
For instance, we have shown that in the case of ice formation of CHLM, a challenging testing ground for the HSEED method which involves
a complex water-substrate interface of relevance for e.g. cryopreservation applications, the HSEED method requires a parallel workload on the order of 10$^5$ CPU hours,
to be compared with the only partially parallelisable 10$^6$ CPU hours needed to converge FFS simulations.
However, it would clearly be desirable to expand the scope of the HSEED approach to non-flat, disordered, rough and flexible
interfaces. This is especially relevant to heterogeneous ice nucleation in biological matter, where most
of the substrates are characterised by complex morphologies that share very little with pristine crystalline surfaces.
The implementation of more sophisticated RSS algorithms could represent a first step in that direction.

The HSEED method could also be used to probe the ice nucleating ability of different nucleation sites within the same crystalline
substrate. This is of paramount importance for e.g. the atmospheric science community, as it is clear 
that the topology of the surface structure of ice nucleating agents such as the mineral feldspar plays a fundamental role
in determining the overall kinetics of ice nucleation~\cite{Kiselev_Science_2016_Fsp100Nuc,
Wang_PCCP_2016_KaoEdgeNucleation}. Thanks to the HSEED method, active sites such as crystalline defects on the nm scales are now within the reach of
atomistic simulations of heterogeneous ice formation. We thus hope that the methodological advancement presented here
will foster a new generation of MD simulations aimed at screening the ice nucleating ability of different compounds,
and so reducing the gap between experiments and simulations. 
Finally, it is worth noticing that the HSEED framework can be extended to include crystallisation scenarios other than
water freezing - thus opening the possibility to accelerate the computational investigation of heterogeneous
nucleation and growth of many other crystalline materials.

\begin{acknowledgments}
This work was supported by the European Research Council under the European Unions
Seventh Framework Programme (FP/2007-2013) / ERC Grant Agreement number 616121 (HeteroIce project). 
We are grateful for computational resources provided by the Materials Chemistry Consortium through the EPSRC grant number EP/L000202 and
the London Centre for Nanotechnology.
\end{acknowledgments}

%\bibliography{Text.bib}
%\bibliographystyle{unsrtnat}
%\end{document}

\end{document}